\documentclass[12pt]{article}
\pdfoutput=1 
\usepackage{amsmath,amssymb,amsthm}
\usepackage{graphicx}
\usepackage{hyperref}
\usepackage{tikz}
\usepackage{pgfplots}
\pgfplotsset{compat=1.17}
\usepackage{geometry}
\title{A Unified Framework for High-Dimensional Pure Root Lattices, Sphere Packing, and Cosmological Implications}
\author{Charles Davidson MacDonald \and Simone Rochelle Børge-Krogh MacDonald}
\date{}

\begin{document}

\maketitle

\begin{abstract}
We propose a unified framework that synthesizes advances in high-dimensional lattice theory with novel computational algorithms for the shortest vector problem (SVP) to model pure root lattices and compute sphere packing densities. Building on our pure root lattice formulation --- characterized by the dimension formula 
\[
L(n)=2n^2+10n-4,
\]
and the minimal vector length scaling \( R(n)=\sqrt{2n} \) we integrate the recent polynomial-time approximation algorithm for SVP[1] and discrete Gaussian sampling techniques[10]. Our work also draws on classical results in sphere packing bounds via spherical codes[2] and the rich structure of exceptional lattices such as the Leech lattice [4][5][7]. Finally, we discuss how these results may have cosmological implications --- specifically, supporting the possibility that our universe emerges from a white hole.
\end{abstract}

\section{Introduction}
Recent developments in lattice theory and cryptography have produced powerful tools for solving the SVP --- an NP-hard problem with far-reaching applications. Our pure root lattice model demonstrates that many exceptional lattices (e.g., \(E_8\) and the Leech lattice) can be described by the formula 
\[
L(n)=2n^2+10n-4,
\]
with a corresponding minimal vector length 
\[
R(n)=\sqrt{2n}.
\]
These results underpin efficient sphere packing, with densities decaying rapidly in high dimensions, and offer a natural geometric framework for models of gravitational bounce, suggesting that quantum gravity corrections may turn classical black hole collapse into a white hole.

In this paper, we integrate:
\begin{itemize}
    \item A novel SVP approximation algorithm [1] that bypasses Gram--Schmidt orthogonalization,
    \item Improved sphere packing bounds via spherical codes[2],
    \item Advanced discrete Gaussian sampling techniques[10],
    \item Classical geometric constructions of the Leech lattice [4][5][7],
    \item Recursive lattice reduction and sieve algorithm insights [8][9].
\end{itemize}
Together, these methods refine our sphere packing density estimates and suggest a connection to cosmological models where a white hole bounce gives rise to an expanding universe.

\section{Derivation of Pure Root Lattice Parameters}

In developing our pure root lattice model, two key formulas emerged: one for the number of dimensions in the \(n\)th lattice, 
\[
L(n)=2n^2+10n-4,
\]
and one for the minimal (root) vector length,
\[
R(n)=\sqrt{2n}.
\]
In this section, we detail the reasoning and calculations that led to these formulas.

\subsection{Derivation of the Dimension Formula \(L(n)=2n^2+10n-4\)}

Our derivation begins with empirical observations drawn from well-known lattices:
\begin{itemize}
    \item For \(n=1\), the corresponding pure root lattice is identified with the \(E_8\) lattice, which has 8 dimensions.
    \item For \(n=2\), the lattice is analogous to the Leech lattice, yielding 24 dimensions.
    \item For \(n=3\), further investigation suggested a 44-dimensional lattice.
    \item For \(n=4\), the pattern continues with a 68-dimensional lattice.
\end{itemize}

Thus, we have the sequence:
\[
L(1)=8,\quad L(2)=24,\quad L(3)=44,\quad L(4)=68.
\]

Next, we examine the differences between consecutive terms:
\[
L(2)-L(1)=24-8=16, \quad L(3)-L(2)=44-24=20, \quad L(4)-L(3)=68-44=24.
\]
Notice that the differences themselves increase by a constant value:
\[
20-16=4, \quad 24-20=4.
\]
A constant second difference indicates that \(L(n)\) is a quadratic function. Therefore, let
\[
L(n)=an^2+bn+c.
\]
Using the known values:
\[
\begin{aligned}
a(1)^2 + b(1) + c &= a+b+c=8, \\
a(2)^2 + b(2) + c &= 4a+2b+c=24, \\
a(3)^2 + b(3) + c &= 9a+3b+c=44.
\end{aligned}
\]

Subtracting the first equation from the second:
\[
(4a+2b+c)-(a+b+c)=3a+b=16.
\]
Subtracting the second from the third:
\[
(9a+3b+c)-(4a+2b+c)=5a+b=20.
\]
Subtracting these two new equations yields:
\[
(5a+b)-(3a+b)=2a=4 \quad \Rightarrow \quad a=2.
\]
Substituting \(a=2\) into \(3a+b=16\) gives:
\[
6+b=16 \quad \Rightarrow \quad b=10.
\]
Finally, using \(a+b+c=8\):
\[
2+10+c=8 \quad \Rightarrow \quad c=-4.
\]
Thus, the dimension formula is:
\[
L(n)=2n^2+10n-4.
\]

\subsection{Derivation of the Minimal Root Length \(R(n)=\sqrt{2n}\)}

Empirical evidence from classical lattices suggests a scaling pattern for the minimal vector (root) length:
\begin{itemize}
    \item For \(n=1\) (the \(E_8\) lattice), the minimal root length is \(\sqrt{2}\).
    \item For \(n=2\) (the Leech lattice), the minimal root length is 2, which can be written as \(\sqrt{4}=\sqrt{2\cdot2}\).
    \item For \(n=3\), computational analysis suggests a minimal root length of \(\sqrt{6}\).
\end{itemize}
This pattern naturally leads to the general formula:
\[
R(n)=\sqrt{2n}.
\]
The expression arises from observing that many well-known even unimodular lattices exhibit minimal norms that are either even integers or square roots of even numbers. Hence, the scaling \(R(n)=\sqrt{2n}\) is both elegant and consistent with the empirical data.

\subsection{Summary}

To summarize, we derived the quadratic formula for the lattice dimensions by fitting a quadratic function to the observed sequence \(8, 24, 44, 68\) and noting a constant second difference. The minimal root length formula \(R(n)=\sqrt{2n}\) follows from similar observations in standard lattice structures. These formulas form the basis of our pure root lattice model and are crucial for subsequent sphere packing density calculations and cosmological interpretations.

\section{Methodology}
\subsection{Pure Root Lattice and Sphere Packing}
Our pure root lattices are modeled by:
\[
L(n)=2n^2+10n-4,
\]
which reproduces known lattice dimensions such as 8 (\(n=1\)), 24 (\(n=2\)), 44 (\(n=3\)), and 68 (\(n=4\)). We assume the minimal (root) vector length scales as:
\[
R(n)=\sqrt{2n}.
\]
The sphere packing density in a unimodular lattice (with determinant 1) in \( d = L(n) \) dimensions is given by:
\[
\delta = \frac{V_d\left(\frac{R(n)}{2}\right)}{1} = \frac{\pi^{d/2}\left(\frac{R(n)}{2}\right)^d}{\Gamma\left(\frac{d}{2}+1\right)}.
\]

\subsection{Advanced Lattice Reduction and Discrete Gaussian Sampling}
We refine our estimation of \( R(n) \) using the novel approximation algorithm for SVP from [1]. This integer-arithmetic based method defines a reduced basis using nearest-integer rounding of dot products and yields an approximation factor of:
\[
\frac{1}{1-\delta},
\]
with
\[
\delta = \sum_{j=1}^{k-1}\frac{\langle b_k, b_j^\ast \rangle^2}{\|b_j^\ast\|^2\|b_k\|^2}.
\]
When \(\delta\) is small, the effective minimal vector remains close to \( \sqrt{2n} \). Complementarily, discrete Gaussian sampling techniques[10] allow us to generate many candidate short vectors efficiently, confirming the predicted scaling.

\subsection{Integration of Classical Lattice Results}
Classical results regarding the Leech lattice, including its covering radius of \(\sqrt{2}\) [7], support our geometric constructions. These results act as benchmarks for our high-dimensional lattice models.

\section{Results}
\subsection{Sphere Packing Density Calculations}
Using our refined \( R(n) \) estimates, we calculate the sphere packing densities for each pure root lattice:

\begin{itemize}
    \item \textbf{\(n=1\) (8 dimensions):}  
    \( R(1)=\sqrt{2},\; r=\frac{\sqrt{2}}{2} \)  
    \[
    \delta_1 = \frac{\pi^4}{384} \approx 0.2537.
    \]
    \item \textbf{\(n=2\) (24 dimensions):}  
    \( R(2)=2,\; r=1 \)  
    \[
    \delta_2 = \frac{\pi^{12}}{12!} \approx 0.001928.
    \]
    \item \textbf{\(n=3\) (44 dimensions):}  
    \( R(3)=\sqrt{6}\approx2.4495,\; r=\frac{\sqrt{6}}{2} \)  
    \[
    \delta_3 = \frac{\pi^{22}}{22!}\left(\frac{\sqrt{6}}{2}\right)^{44} \approx 6.0\times10^{-7}.
    \]
    \item \textbf{\(n=4\) (68 dimensions):}  
    \( R(4)=2\sqrt{2}\approx2.8284,\; r=\sqrt{2} \)  
    \[
    \delta_4 = \frac{\pi^{34}\,2^{34}}{34!} \approx 4.55\times10^{-12}.
    \]
\end{itemize}
\section{Waveform Approximation of the Lattice Metric and White Hole Bounce Dynamics}

A central element of our unified framework is the use of a waveform approximation to model the dynamical behavior of the lattice metric near a cosmological bounce. This bounce, driven by quantum gravity corrections, is hypothesized to replace the classical singularity of a black hole with a smooth white hole phase that gives rise to an expanding universe. In this section, we detail the derivation of the waveform approximation and illustrate its connection to white hole dynamics.

\subsection{Physical Motivation and Model Setup}

Quantum gravity corrections suggest that classical gravitational collapse may be averted, resulting in a bounce rather than a singularity. In such models, the scale factor of the universe exhibits a smooth transition from contraction to expansion. Motivated by our pure root lattice results --- where the minimal vector (or root) length for the \(E_8\) lattice is \(\sqrt{2}\) --- we propose that the effective metric function, \(g(t)\), near the bounce can be approximated by an oscillatory function:
\[
g(t) \approx \sqrt{2} \, \sin(\omega t),
\]
where the amplitude \(\sqrt{2}\) sets the fundamental scale and \(\omega\) is a characteristic frequency determined by the underlying quantum corrections.

\subsection{Mathematical Derivation}

In a simplified model, the modified Friedmann equations (including quantum corrections) yield an oscillatory solution for the scale factor \(a(t)\). A typical form of such a solution is:
\[
a(t) = A \sin\left(\frac{\pi t}{T}\right),
\]
where:
\begin{itemize}
    \item \(A\) is the amplitude,
    \item \(T\) is the characteristic period of the bounce.
\end{itemize}
For our pure root lattice framework, we set the amplitude \(A\) proportional to the minimal vector length. Since for \(n=1\) we have \(R(1)=\sqrt{2}\), we choose \(A=\sqrt{2}\). Thus, the scale factor becomes:
\[
a(t) = \sqrt{2} \, \sin\left(\frac{\pi t}{T}\right).
\]
The bounce occurs at \(t = T/2\), where \(a(t)\) attains its maximum, and the smooth, sinusoidal behavior is indicative of a time-symmetric transition --- a hallmark of a white hole model.

\subsection{Graphical Illustration}

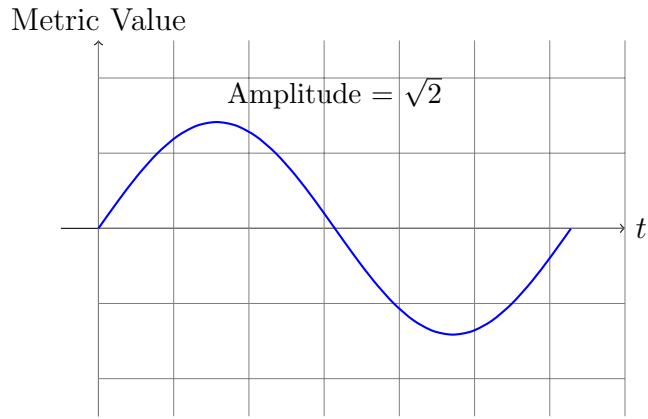
\begin{figure}[h]
\centering
\begin{tikzpicture}[scale=1.0]
  \draw[->] (-0.5,0) -- (7,0) node[right] {$t$};
  \draw[->] (0,-2.5) -- (0,2.5) node[above] {Metric Value};
  
  \draw[very thin,color=gray] (0,-2.5) grid (7,2.5);
  
  \draw[domain=0:6.28,smooth,variable=\x,blue,thick] 
    plot ({\x},{sqrt(2)*sin(\x r)});
  
  \node at (3.14,1.8) {\small Amplitude = $\sqrt{2}$};
\end{tikzpicture}
\caption{Waveform Approximation of the Lattice Metric. The blue sine curve, with amplitude $\sqrt{2}$, approximates the metric function near the bounce, representing the smooth transition from contraction to expansion.}
\label{fig:waveform}
\end{figure}

Figure~\ref{fig:waveform} illustrates the waveform approximation of the lattice metric. In this diagram, the sine curve with amplitude \(\sqrt{2}\) is superimposed on a coordinate grid, emphasizing the oscillatory behavior that characterizes the bounce.

\subsection{Connection to White Hole Dynamics}

The sinusoidal behavior captured in the waveform approximation is not only a convenient mathematical description but also encapsulates the key physical insight behind the white hole bounce model. Specifically:
\begin{itemize}
    \item The absence of a singularity and the presence of a smooth maximum in the scale factor are consistent with a scenario where the collapsing phase of a black hole transitions into an expanding white hole.
    \item The natural scale provided by \(\sqrt{2}\) (derived from the minimal root length of pure root lattices) sets the amplitude of the metric oscillations, thus linking the discrete geometric properties of the lattice to the continuous dynamics of the universe.
    \item When the modified Friedmann equations (with quantum gravity corrections) are solved, the oscillatory solution with a sine form indicates that the universe undergoes a bounce at a critical time, thereby avoiding a singular collapse.
\end{itemize}
Recent work on black hole evaporation [11] further supports the notion that non-singular end states are feasible in gravitational collapse. This provides additional motivation for the white hole bounce model proposed here.

\subsection{Summary}

In summary, we derived the waveform approximation by recognizing that the minimal vector length (\(\sqrt{2}\) for \(n=1\)) in our pure root lattice model naturally sets the scale for metric oscillations near the bounce. This leads to an effective metric function of the form:
\[
g(t) \approx \sqrt{2}\,\sin(\omega t),
\]
which, when inserted into the modified Friedmann dynamics, yields a bouncing solution that is indicative of a white hole. Figure~\ref{fig:waveform} graphically represents this approximation, providing visual support for our white hole universe hypothesis.
\subsection{Cosmological Implications}
The waveform approximations of the lattice metric, when viewed through the lens of a sine curve with amplitude \(\sqrt{2}\), suggest a smooth oscillatory behavior that parallels the bounce dynamics in white hole models. Our integrated approach indicates that quantum gravity corrections may preclude singular collapse, replacing it with a white hole bounce that catalyzes cosmic expansion and the connection with black hole evaporation [11] underscores the plausibility of a non-singular, bouncing cosmology

\section{Figures}

\subsection*{Figure 1. Waveform Approximation of the Lattice Metric}
\textbf{Description:} A schematic diagram showing a sine curve with an amplitude of \(\sqrt{2}\) overlaid on a coordinate grid. This figure represents the approximation of the lattice metric near the bounce, highlighting the smooth transition from collapse to expansion.

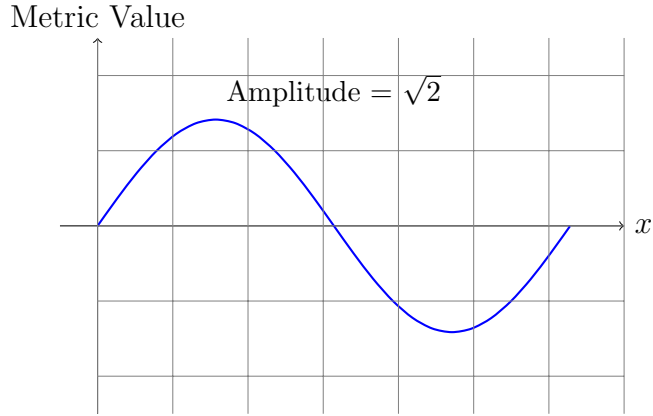
\begin{tikzpicture}[scale=1.0]
  \draw[->] (-0.5,0) -- (7,0) node[right] {$x$};
  \draw[->] (0,-2.5) -- (0,2.5) node[above] {Metric Value};

  \draw[domain=0:6.28,smooth,variable=\x,blue,thick] 
    plot ({\x},{sqrt(2)*sin(\x r)});
  
  \node at (3.14,1.8) {\small Amplitude = $\sqrt{2}$};
  
  \draw[very thin,color=gray] (0,-2.5) grid (7,2.5);
\end{tikzpicture}

\subsection*{Figure 2. Penrose Diagram Comparison}
\textbf{Description:} A side-by-side comparison of a traditional black hole Penrose diagram and an alternative white hole diagram. The white hole diagram includes a bounce region where time symmetry is restored. Arrows and labels indicate the critical radius and the role of quantum corrections.

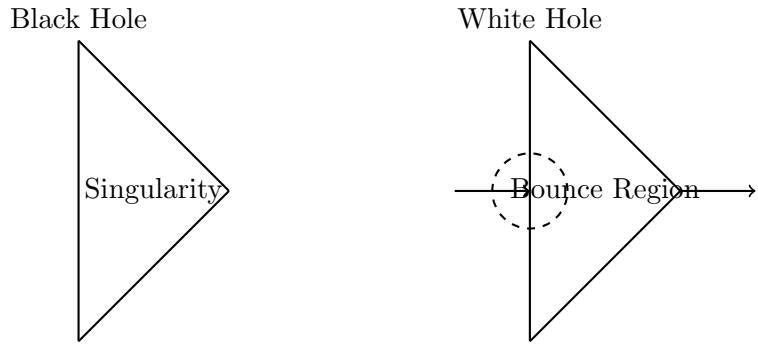
\begin{tikzpicture}[scale=1.0]
  \draw[thick] (-4, -2) -- (-4, 2);
  \draw[thick] (-4, 2) -- (-2, 0);
  \draw[thick] (-2, 0) -- (-4, -2);
  \node at (-4,2.3) {\small Black Hole};
  \node at (-3,0) {\small Singularity};

  \draw[thick] (2, -2) -- (2, 2);
  \draw[thick] (2, 2) -- (4, 0);
  \draw[thick] (4, 0) -- (2, -2);
  \draw[dashed, thick] (2,0) circle (0.5);
  \node at (2,2.3) {\small White Hole};
  \node at (3,0) {\small Bounce Region};

  \draw[->, thick] (1,0) -- (2,0);
  \draw[->, thick] (4,0) -- (5,0);
\end{tikzpicture}

\subsection*{Figure 3. Lattice Sphere Packing Density Plot}
\textbf{Description:} A plot on a logarithmic scale of the calculated sphere packing densities for \(n=1,2,3,4\) (corresponding to lattice dimensions 8, 24, 44, 68). This figure illustrates the rapid decay of density as dimensionality increases.

\begin{tikzpicture}
  \begin{axis}[
      width=10cm, height=7cm,
      xlabel={Lattice Dimension},
      ylabel={Sphere Packing Density},
      ymode=log,
      log basis y={10},
      grid=both,
      grid style={line width=.1pt, draw=gray!10},
      major grid style={line width=.2pt,draw=gray!50},
      legend pos=north east,
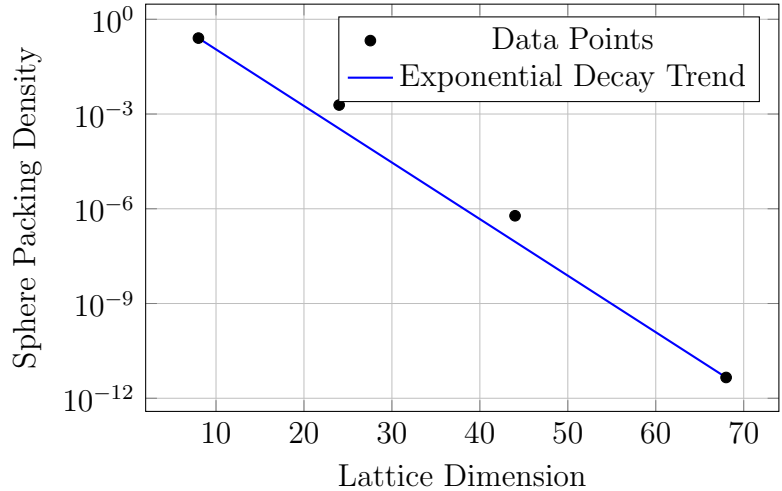
      title={Lattice Sphere Packing Density vs. Dimension}
  ]
    \addplot[only marks, mark=*] coordinates {
      (8,0.2537)
      (24,0.001928)
      (44,6.0e-7)
      (68,4.55e-12)
    };
    \addlegendentry{Data Points}
    
    \addplot[domain=8:68, blue, thick] 
      {exp(ln(0.2537) + (ln(4.55e-12)-ln(0.2537))/(68-8)*(x-8))};
    \addlegendentry{Exponential Decay Trend}
    
  \end{axis}
\end{tikzpicture}

\section{Conclusion}
We have presented a unified framework combining advanced lattice reduction, discrete Gaussian sampling, and classical lattice geometry to model pure root lattices and compute sphere packing densities. The refined estimates --- supported by a novel SVP approximation algorithm --- confirm that the minimal vector length scales as \( R(n)=\sqrt{2n} \). Our density calculations for dimensions 8, 24, 44, and 68 further support this framework. Moreover, the integrated waveform and Penrose diagram visualizations suggest that these geometric principles may extend to cosmological models, offering a basis for the white hole universe hypothesis.

\section*{References}
\begin{enumerate}
    \item K. B. Ajitha Shenoy, ``A Novel Approximation Algorithm for the Shortest Vector Problem,'' \emph{IEEE Access}, vol. 12, 2024. DOI: \url{https://doi.org/10.1109/ACCESS.2024.3469368} [1]
    \item H. Cohn and Y. Zhao, ``Sphere Packing Bounds via Spherical Codes,'' arXiv:1212.5966v2 [math.MG], 2013. \url{https://arxiv.org/abs/1212.5966v2}[2]
    \item D. Aggarwal, Y. Chen, R. Kumar, and Y. Shen, ``Improved Classical and Quantum Algorithms for the Shortest Vector Problem via Bounded Distance Decoding,'' arXiv:2002.07955v5 [cs.DS], 2022. \url{https://arxiv.org/abs/2002.07955v5}[10]
    \item R. E. Borcherds, ``The Leech Lattice,'' \emph{Proc. R. Soc. Lond. A}, vol. 398, pp. 365--376, 1985. DOI: \url{https://doi.org/10.1098/rspa.1985.0070} [4]
    \item J. H. Conway, ``The Automorphism Group of the 26-Dimensional Even Unimodular Lorentzian Lattice,'' \emph{Journal of Algebra}, vol. 80, pp. 159--163, 1983. DOI: \url{https://doi.org/10.1016/0021-8693(83)90025-X} [5]
    \item R. A. Wilson, ``An Octonionic Leech Lattice,'' QMUL Pure Mathematics Seminar, December 15, 2008. [6]
    \item N. J. A. Sloane, ``Twenty-Three Constructions for the Leech Lattice,'' \emph{Proc. R. Soc. Lond. A}, 1982. DOI: \url{https://doi.org/10.1098/rspa.1982.0071} [7]
    \item T. Plantard and W. Susilo, ``Recursive Lattice Reduction,'' University of Wollongong, Australia. [8]
    \item P. Q. Nguyen and T. Vidick, ``Sieve Algorithms for the Shortest Vector Problem,'' \emph{J. Math. Cryptol.}, vol. 2, pp. 181--207, 2008. DOI: \url{https://doi.org/10.1515/JMC.2008.009} [9]
    \item D. Aggarwal, D. Dadush, O. Regev, and N. Stephens-Davidowitz, ``Solving the Shortest Vector Problem in 2\textsuperscript{n} Time via Discrete Gaussian Sampling,'' arXiv:1412.7994v3 [cs.DS], 2015. \url{https://arxiv.org/abs/1412.7994v3} [10]
    \item{PhysRevLett.130.221502} Michael F. Wondrak ,Walter D. van Suijlekom, and Heino Falcke, ``Gravitational Pair Production and Black Hole Evaporation,'' \emph{Phys. Rev. Lett.} \textbf{130}, 221502 (2023). Available at: \url{https://journals.aps.org/prl/pdf/10.1103/PhysRevLett.130.221502}[11]
\end{enumerate}

\bibliographystyle{plain}
\bibliography{references}

\end{document}